\documentclass[11pt]{article}
\usepackage[final]{acl}
\usepackage{times}
\usepackage{booktabs}
\usepackage{amsmath}
\usepackage{dsfont}
\usepackage{graphicx}
\usepackage[T1]{fontenc}
\usepackage[utf8]{inputenc}
\usepackage{xspace}
\usepackage{xcolor}
\IfFileExists{newunicodechar.sty}{%
\usepackage{newunicodechar}%
\newunicodechar{ć}{\'c}%
\newunicodechar{š}{\v{s}}}{}
\definecolor{speech}{HTML}{2E68DB}


\newcommand{\AvailableLangs}{6\xspace}
\newcommand{\PairHiCer}{15.25\xspace}
\newcommand{\PairFrCer}{15.26\xspace}
\newcommand{\PairHiAnsPct}{3.6\%\xspace}
\newcommand{\PairFrAnsPct}{79.3\%\xspace}
\newcommand{\YesnoBalLoPct}{45.5\%\xspace}
\newcommand{\YesnoBalHiPct}{54.0\%\xspace}
\newcommand{\YesnoBalMeanPct}{48.6\%\xspace}
\newcommand{\YesnoScoredLangs}{7\xspace}
\newcommand{\YesnoMinScored}{100\xspace}
\newcommand{\YesnoChancePct}{50\%\xspace}
\newcommand{\AvailableAtPct}{50\%\xspace}
\newcommand{\TopicChancePct}{14.3\%\xspace}
\newcommand{\NTopics}{7\xspace}
\newcommand{\TopicNumAnsLoPct}{26.0\%\xspace}
\newcommand{\TopicNumAnsHiPct}{100.0\%\xspace}
\newcommand{\TopicMlAccPct}{10.9\%\xspace}
\newcommand{\TopicNumAccPct}{16.0\%\xspace}
\newcommand{\TopicMajorityPct}{30.6\%\xspace}
\newcommand{\TopicNumBalPct}{23.2\%\xspace}
\newcommand{\TopicMlAnsPct}{65.0\%\xspace}
\newcommand{\TopicOpenAnsHiPct}{16.0\%\xspace}
\newcommand{\RefBelebeleSeconds}{47.0\xspace}
\newcommand{\RefQwenAudioHours}{520k\xspace}
\newcommand{\AnswerMaxExtraWords}{8\xspace}
\newcommand{\QwenAudioBalPct}{39.4\%\xspace}
\newcommand{\QwenAudioBelowMaj}{4\xspace}
\newcommand{\QwenBelowMajSilent}{3\xspace}
\newcommand{\NLangs}{11\xspace}
\newcommand{\TextTopicBalPct}{71.5\%\xspace}
\newcommand{\SpeechTopicBalPct}{23.2\%\xspace}
\newcommand{\TextTopicAnsPct}{100.0\%\xspace}
\newcommand{\TextYesnoBalPct}{49.2\%\xspace}
\newcommand{\TextYesnoAnsLoPct}{25.7\%\xspace}
\newcommand{\TextYesnoAnsHiPct}{98.5\%\xspace}
\newcommand{\TextYesnoBalLoPct}{45.0\%\xspace}
\newcommand{\TextYesnoBalHiPct}{54.4\%\xspace}
\newcommand{\YesnoIfrMean}{49.8\xspace}
\newcommand{\TopicIfrMean}{81.3\xspace}
\newcommand{\TextRecallMean}{71.5\xspace}
\newcommand{\DecBaseSpeechPct}{14.8\%\xspace}
\newcommand{\DecBaseTextPct}{17.4\%\xspace}
\newcommand{\DecTextLoPct}{71.4\%\xspace}
\newcommand{\DecTextHiPct}{73.8\%\xspace}
\newcommand{\DecRegionalLoPct}{26.2\%\xspace}
\newcommand{\DecRegionalHiPct}{27.3\%\xspace}
\newcommand{\DecQwenSpeechPct}{25.5\%\xspace}
\newcommand{\DecQwenTextPct}{73.3\%\xspace}
\newcommand{\DecQwenVsGlobalPp}{2.3\xspace}
\newcommand{\DecMatchGainPp}{3.7\xspace}
\newcommand{\EarthLangs}{Amharic, Hausa and Swahili\xspace}
\newcommand{\FireLangs}{Hindi, Marathi, Tamil and Urdu\xspace}
\newcommand{\WaterLangs}{English, French, Indonesian and Spanish\xspace}
\newcommand{\SluCascadeRecallPct}{52.5\%\xspace}
\newcommand{\SluCascadeAnsPct}{100.0\%\xspace}
\newcommand{\TeacherPersonPct}{13.1\%\xspace}
\newcommand{\HoursHiIn}{2,189\xspace}
\newcommand{\HoursUrPk}{81\xspace}
\newcommand{\UrAnsPct}{65.6\%\xspace}
\newcommand{\TopicOpenAccHiPct}{91\%\xspace}
\newcommand{\TopicOpenAnsHiIntPct}{16\%\xspace}
\newcommand{\MlAnsFewerN}{8\xspace}
\newcommand{\SibTextAccPct}{88.9\%\xspace}
\newcommand{\SibCascadeAccPct}{86.4\%\xspace}
\newcommand{\SibSpeechLoPct}{21.3\%\xspace}
\newcommand{\SibSpeechHiPct}{80.9\%\xspace}
\newcommand{\TrainSteps}{1,000\xspace}
\newcommand{\TrainEffectiveBatch}{512\xspace}
\newcommand{\TrainOptim}{AdamW\xspace}
\newcommand{\TrainLr}{$1.5\times10^{-3}$\xspace}
\newcommand{\TrainLrQwen}{$1.0\times10^{-3}$\xspace}
\newcommand{\TrainSchedule}{cosine\xspace}
\newcommand{\TrainWarmup}{10\xspace}
\newcommand{\TrainMaxSec}{30\xspace}
\newcommand{\TrainRuns}{66\xspace}
\newcommand{\TrainWallMedianH}{15.2\xspace}
\newcommand{\TinyAyaParamsB}{3.66\xspace}
\newcommand{\QwenParamsB}{4.33\xspace}
\newcommand{\ProjParamsTinyAyaM}{2.1\xspace}
\newcommand{\ProjParamsQwenM}{2.6\xspace}

\title{Zero-Shot Spoken Language Understanding with SLAM-ASR:\\
What Do Monolingual ASR-Trained Linear Projectors Enable?}

\author{Edwin Arkel Rios$^{1,2}$\thanks{Work done during an internship at Tranxform AI.}, Sneh Sourabh$^{3}$, Jocelyn Zaruma$^{4,5}$, Girish Ewoorkar$^{6}$, \\
\textbf{Julian Mack}$^{7}$, \textbf{Bo-Cheng Lai}$^{1}$ \\
$^{1}$National Yang Ming Chiao Tung University, Taiwan \quad $^{2}$Tranxform AI \\
$^{3}$Indian Institute of Technology Hyderabad, India \\
$^{4}$Institut national des langues et civilisations orientales, France \\
$^{5}$Université Sorbonne Nouvelle, France \quad $^{6}$Cohere Labs Community \quad $^{7}$Cohere}

\begin{document}
\maketitle

\begin{abstract}
The SLAM-ASR framework connects the features of a frozen speech encoder as input to a frozen LLM decoder to enable speech transcription. In this work we ask whether this training---on speech transcription---provides the LLM with the capability to also perform spoken language understanding (SLU) tasks, by training monolingual projectors on \NLangs{} languages. 
When prompted to answer a yes/no question about the input speech, the mean instruction following rate (IFR) is \YesnoIfrMean{}\%. When not following the instruction, these models default back to transcription. The transcription quality---measured by the character error rate (CER)---is not sufficient to predict IFR. 
Remarkably, when the model is instructed to answer the topic question as a single digit option, the mean IFR rises to \TopicIfrMean{}\%. Our models score a mean \TopicNumBalPct{} topic balanced accuracy (BAcc) against \TopicChancePct{} chance.
However, the same decoder scores a mean \TextRecallMean{}\% BAcc when using reference transcripts as text input. This indicates a failure of the projector to transfer information for the SLU tasks.
\end{abstract}

\section{Introduction}
The SLAM-ASR recipe \citep{ma_embarrassingly_2024} enables training of large audio language models (LALMs) by connecting a speech encoder to a text decoder through a linear projector. LALMs enable large language models (LLMs) to perform spoken language understanding (SLU), but most works following the SLAM-ASR recipe have been limited to studying ASR capabilities \cite{fong_speech_2025,kumar_performance_2025,zhang_language_2026,burdisso_reducing_2026}.
Few works have explored zero-shot SLU---translation \cite{fan_alignformer_2025} and intent classification \cite{lee_towards_2026}---for LALMs beyond transcription, using a trainable transformer-based projector instead. However, this increases the cost of the system.
In this study, we ask whether a linear projector---trained on monolingual ASR data---can enable SLU capabilities for an LLM, and if yes, under which conditions.

\section{Setup}
We follow the SLAM-ASR recipe \citep{ma_embarrassingly_2024}. A Whisper-medium speech encoder \citep{radford_robust_2022} is connected to a Tiny Aya Global decoder \citep{salamanca_tiny_2026} via a linear projector, while the encoder and decoder are kept frozen.
We train a projector for each of the \NLangs{} target languages using WorldSpeech \citep{asonitis_worldspeech_2026} with an audio transcription loss. % on an ASR objective
We compare our models against a strong baseline, Qwen2-Audio-7B \citep{chu_qwen2-audio_2024}, a \emph{multi-task LALM}. It also connects a Whisper encoder to an LLM decoder through a linear projector, but it is trained with supervision on \RefQwenAudioHours{} hours of audio data, across many tasks, each posed as a natural-language instruction. Training details are in Appendix~\ref{sec:training}.

For evaluation, we use FLEURS \citep{conneau_fleurs_2023}. We evaluate three tasks: 1) transcription, 2) yes/no questions, and 3) topic classification. We use character error rate (CER) \cite{k_advocating_2025} to evaluate transcription quality. For yes/no questions and topic classification we measure instruction following rate (IFR), the share of utterances whose output counts as a valid answer. For yes/no questions we also use balanced accuracy (BAcc), the mean of each class's recall over answered utterances.
A teacher model uses reference transcripts to create proxy labels for the yes/no task---whether a person is being named in the script/audio or not. For the topic classification task, we measure BAcc, using the labels from SIB-Fleurs \citep{schmidt_fleurs-slu_2025}, which classifies samples into \NTopics{} categories.
We adopt its English prompt template, which we call the \emph{digit template} (topics use digits as choice labels). An audio output counts as a valid answer only if it names exactly one option with at most \AnswerMaxExtraWords{} other words. A text output counts as a valid answer if it names any option, and the first one named is its answer. Details on metrics are in Appendix~\ref{sec:metrics}.

% GENERATED by build_slu_abstract_tables.py -- edit that script, not this file.

\begin{table}[t]
\centering
\footnotesize
\setlength{\tabcolsep}{4pt}
\begin{tabular}{lrrrrr}
\toprule
 & & & \multicolumn{3}{c}{Topic Classification} \\
\cmidrule(lr){4-6}
 & & Yes/no & \multicolumn{2}{c}{Speech} & Text \\
\cmidrule(lr){4-5}
Language & CER\% & IFR\% & IFR\% & BAcc\% & BAcc\% \\
\midrule
English & 6.65 & 93.5 & 100.0 & 41.4 & 73.0 \\
Spanish & 3.09 & 86.9 & 100.0 & 36.2 & 73.6 \\
French & 15.26 & 79.3 & 99.4 & 35.7 & 75.8 \\
Indonesian & 6.35 & 91.0 & 100.0 & 35.7 & 71.3 \\
Swahili & 21.61 & 92.8 & 99.6 & 17.1 & 66.0 \\
Urdu & 28.67 & 65.6 & 100.0 & 16.4 & 70.9 \\
Amharic & 39.86 & 23.3 & 79.6 & 14.7 & 70.4 \\
Hausa & 57.05 & 1.8 & 74.3 & 14.9 & 67.2 \\
Hindi & 15.25 & 3.6 & 26.0 & 14.3 & 71.5 \\
Marathi & 19.11 & 9.5 & 41.4 & 14.3 & 71.1 \\
Tamil & 87.87 & 0.2 & 73.6 & 14.3 & 75.2 \\
\bottomrule
\end{tabular}
\caption{Performance across languages. CER measures transcription quality. IFR is the instruction following rate. Topic Classification uses the digit template and reports balanced accuracy (BAcc) across 7 topics for Speech (audio via the projector) and Text (reference transcript replaces the audio).}
\label{tab:perlang}
\end{table}

\section{Results}
\paragraph{Can a projector trained only for transcription answer questions? Yes.} Following a prompt for a task other than transcription suggests some instruction following. For \AvailableLangs{} of \NLangs{} projectors, more than half of yes/no outputs are a short answer rather than a transcript (Table~\ref{tab:perlang}). Recognition quality alone does not decide which projectors answer. Despite nearly identical CERs for Hindi (\PairHiCer{}) and French (\PairFrCer{}), their IFRs diverge drastically (\PairHiAnsPct{} vs. \PairFrAnsPct{}). When a projector fails to answer, it reverts to transcription. 

\paragraph{Can a projector answer correctly?}
Answering correctly would indicate understanding of the utterance's content. Evaluated against teacher labels, yes/no BAcc shows our models are not better than chance (\YesnoBalLoPct{}--\YesnoBalHiPct{} in the \YesnoScoredLangs{} languages with at least \YesnoMinScored{} answers). Open-ended topic queries phrased in the projector's language elicit answers on at most \TopicOpenAnsHiPct{} of utterances. Remarkably, the digit template increases IFR to \TopicNumAnsLoPct{}--\TopicNumAnsHiPct{} (Table~\ref{tab:perlang}). Raw accuracy averages \TopicNumAccPct{}, below the
\TopicMajorityPct{} obtained by always predicting the most frequent topic. BAcc averages \TopicNumBalPct{} over the \NLangs{} languages against \TopicChancePct{}. The translated digit template lowers IFR to \TopicMlAnsPct{}; its per-language results are in Appendix~\ref{sec:appendix}. On identical utterances and answer rules, the multi-task LALM achieves a higher mean BAcc (\QwenAudioBalPct{}) than our projectors. Yet on \QwenAudioBelowMaj{} of \NLangs{} languages---\QwenBelowMajSilent{} where our projectors answer few yes/no questions---its raw accuracy falls below that of always predicting the most frequent topic. These languages are difficult even for a multi-task LALM.

\paragraph{Does the limitation come from the projector or the decoder?}
We evaluate the decoder alone on each question using the
FLEURS reference text as input in place of the audio. With the text input, the decoder answers
\TextTopicAnsPct{} of topic questions at a mean \TextTopicBalPct{} BAcc,
against \SpeechTopicBalPct{} through its own projector with audio input. This gap holds in every language. The text-only version is nearly the same across languages while CER varies widely, so the gap does not track the decoder's per-language competence (Table~\ref{tab:perlang}). These BAcc values count only answered utterances. The projector answers fewer utterances than the text input yields, so scoring unanswered utterances as wrong would widen the gap further. This speech-text ceiling gap aligns with the findings of
\citet{cuervo_closing_2026}, attributing the issue to the audio-text projector. The yes/no task stays at chance for both audio (\YesnoBalMeanPct{} BAcc) and text (\TextYesnoBalPct{}). Detailed results are in Appendix~\ref{sec:appendix}.

\section{Conclusion}
Instruction following emerges without any labeled task example. For over half the evaluated languages, the models answer in the right form. However, the answers show limited understanding of the utterance. Yes/no BAcc stays at chance, and topic BAcc stays far below what the same decoder reaches from the reference transcript. Table~\ref{tab:perlang} outlines which languages follow instructions for SLU tasks. When not following instructions, these models default to transcription. Notably, topic BAcc exceeds double the \TopicChancePct{} chance only in English, Spanish, French, and Indonesian. Because Swahili and Urdu yield high response rates with near-chance BAcc, non-answering projectors present lower deployment risk than randomly answering ones. The text ceiling demonstrates that the topic knowledge exists in the decoder; what fails is the projector's alignment to it.

\paragraph{Limitations.}
All projectors use one speech encoder, Whisper-medium, and each projector was trained once, with a single random seed. The teacher labels are not independently validated and are generated from the source evaluation dataset's reference transcripts. They measure what is said, not how it is said, and that may affect the decoder output when speech is the input. FLEURS consists of read speech from Wikipedia sentences, which is narrow in domain. We used only the utterance-level samples of Fleurs-SLU because its listening-comprehension samples average \RefBelebeleSeconds{}~s, well beyond the Whisper-medium \TrainMaxSec{}~s window used in our study.
The in-language templates (Table~\ref{tab:prompts-ml}) are LLM translations of the English template, without native-speaker review.
We did not test sensitivity to the template format, for example digits against letters as choice labels, or different delimiters.
Under the audio answer rule, a correct option inside a longer output counts as unanswered.
Answering behavior could improve with further training on data beyond simple transcription.

\section*{Acknowledgments}
This work was supported by Tranxform AI and NSTC under Grant 114-2221-E-A49-186.

\bibliography{refs}

\raggedbottom
\appendix
\section{Training Setup}
\label{sec:training}
The main text uses Tiny Aya Global as the decoder, the globally balanced instruction-tuned variant. Appendix~\ref{sec:appendix} applies the same setup to the other four Tiny Aya variants: Base, the model before instruction tuning, and the region-specialized variants Earth (Africa and West Asia: \EarthLangs{}), Fire (South Asia: \FireLangs{}) and Water (Asia-Pacific and Europe: \WaterLangs{}), each of our languages listed under the variant whose region the Tiny Aya report assigns it. Together with Qwen3-4B, six decoders for each of the \NLangs{} languages give \TrainRuns{} projectors. Each model has three parts: a frozen Whisper-medium encoder, a linear projector, and a frozen decoder. Only the projector is trained, on one language's WorldSpeech transcription data, with samples of at most \TrainMaxSec{}~s. The projector has \ProjParamsTinyAyaM{}M parameters with a Tiny Aya decoder and \ProjParamsQwenM{}M with Qwen3-4B. The encoder, projector and decoder together have \TinyAyaParamsB{}B parameters with any Tiny Aya decoder and \QwenParamsB{}B with Qwen3-4B. Training runs \TrainSteps{} steps at an effective batch of \TrainEffectiveBatch{} samples, with \TrainOptim{} on a \TrainSchedule{} schedule and \TrainWarmup{} warmup steps. The learning rate is \TrainLr{} for Tiny Aya and \TrainLrQwen{} for Qwen3-4B. We evaluate the checkpoint the trainer recorded as best on validation loss, not the final one. A run takes a median \TrainWallMedianH{}~h on one H200 GPU.

\section{Evaluation Details}
\label{sec:evaluation}
\subsection{Metrics}
\label{sec:metrics}
For evaluation, each question asks for an answer from a fixed set of options. An option is named by its digit in the digit template, by a topic name in the prompt's language in the open question, and by a yes or no word of the prompt's language in the yes/no question. With audio input, an answering model writes a bare option, such as ``5'' for the digit template. A model that does not answer usually writes a transcript, which can name an option by chance. With text input, the decoder reads the reference transcript instead of the audio, or the projector's own transcript in the cascade of Appendix~\ref{sec:appendix}. The decoder given text names an option and then explains it, as in ``5 the utterance is about science because it discusses chemistry''.

Each output is first lowercased and stripped of diacritics, Indic vowel signs, bracketed text and punctuation. Prompt echoes are removed, and immediately repeated phrases are kept once. Words are the tokens that remain, so an Indic word can count as several. With audio input, an output counts as an answer only if it names exactly one option and has at most \AnswerMaxExtraWords{} other words. With text input, an output that names any option counts as an answer, and the first option named is the answer.

Labelled utterances are every recording for audio input and each unique FLEURS sentence for text input. Let $N$ be the labelled utterances, $A \subseteq N$ the answered ones, $\hat y_i$ the answer and $y_i$ the label. The instruction following rate (IFR), accuracy (Acc.), accuracy over answered utterances (Acc.$^\dagger$) and balanced accuracy (BAcc) are
\begin{align*}
\text{IFR} &= |A| / |N|, \\
\text{Acc.} &= \textstyle\frac{1}{|N|}\sum_{i \in A} \mathds{1}[\hat y_i = y_i], \\
\text{Acc.}^\dagger &= \textstyle\frac{1}{|A|}\sum_{i \in A} \mathds{1}[\hat y_i = y_i], \\
\text{BAcc} &=\frac{1}{|C|}\sum_{c \in C} \frac{|\{i \in A : y_i = \hat y_i = c\}|}{|\{i \in A : y_i = c\}|},
\end{align*}
where $\mathds{1}[\cdot]$ is 1 when the condition inside holds and 0 otherwise. $C$ is the set of labels among the answered utterances: yes and no for the yes/no task, and the topics for the topic classification task. $C$ contains every class in every reported cell except one---Tiny Aya Base on Marathi speech (Table~\ref{tab:decoders})---where the few answered utterances do not cover every topic. For yes/no IFR, $N$ contains every test utterance, labelled or not.

\subsection{Prompts and Labels}
\label{sec:prompts}
% GENERATED by build_slu_abstract_tables.py -- edit that script, not this file.

\begin{table*}[t]
\centering
\small
\begin{tabular}{@{}p{0.24\textwidth}p{0.73\textwidth}@{}}
\toprule
Question & Prompt \\
\midrule
Yes/no, person named & Answer with one word, yes or no: does the audio mention a person by name? \textcolor{speech}{\textbf{\{AUDIO\}}} Answer: \\
Topic, open question & What is the topic of the audio? Answer with one or two words. \textcolor{speech}{\textbf{\{AUDIO\}}} Topic: \\
Topic, digit template & The utterance belongs to one of the following topics. Utterance: \textcolor{speech}{\textbf{\{AUDIO\}}} Topics: 1. Entertainment 2. Geography 3. Health 4. Politics 5. Science and Technology 6. Sports 7. Travel Please respond with only the number of the correct answer (1, 2, 3, 4, 5, 6, or 7). Answer: \\
\bottomrule
\end{tabular}
\caption{The English prompts as the models saw them. \textcolor{speech}{\textbf{\{AUDIO\}}} marks the position of the projected audio tokens, and the model continues after the final colon. Table~\ref{tab:prompts-ml} gives the in-language template for each language.}
\label{tab:prompts}
\end{table*}

% GENERATED by build_slu_abstract_tables.py -- edit that script, not this file.

\begin{table*}[t]
\centering
\includegraphics[width=\textwidth]{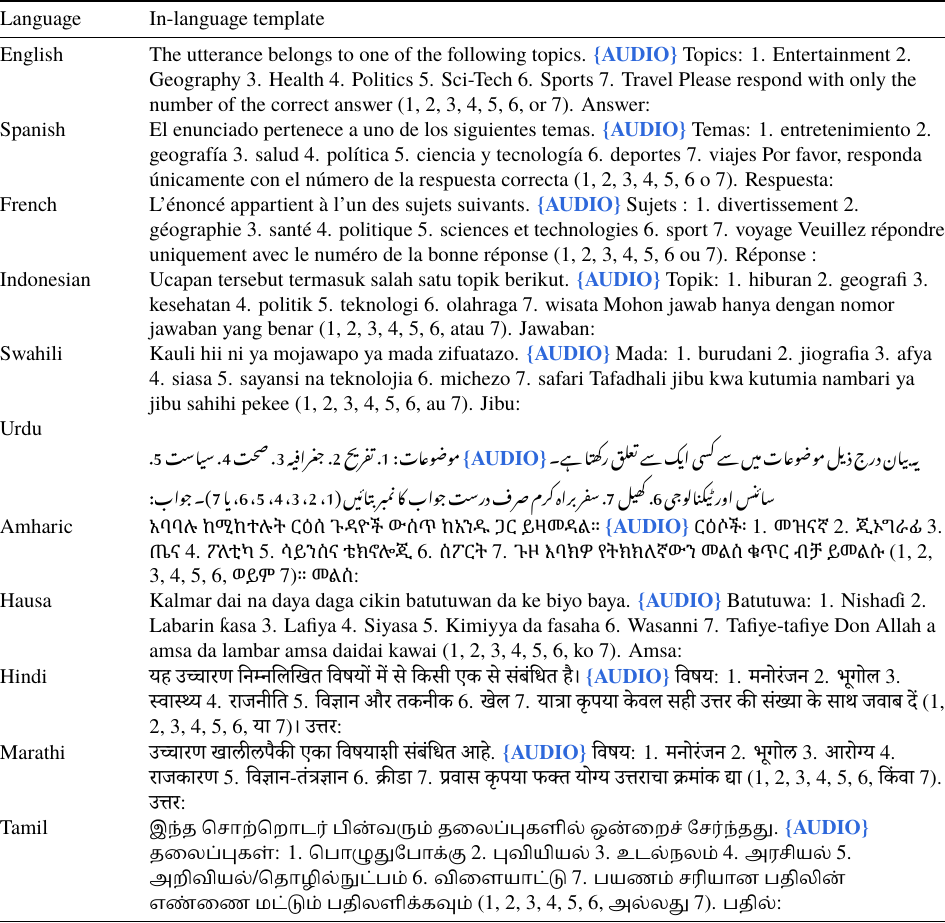}
\caption{The in-language digit template for each language, as the models saw it. It translates the English template\textquotesingle s wrapper and option labels into the projector\textquotesingle s language and keeps the digit answer space. Its English row drops ``Utterance:'' and shortens a label, so it differs from the digit template of Table~\ref{tab:prompts} used in Table~\ref{tab:perlang}.}
\label{tab:prompts-ml}
\end{table*}

Table~\ref{tab:prompts} gives the English prompts as the models saw them, and Table~\ref{tab:prompts-ml} gives the in-language template for each language. The yes/no labels come from a text-only teacher: Claude Sonnet reads the English side of each FLEURS parallel sentence and states whether it names a person, and the label carries to every language because the sentences are parallel. \TeacherPersonPct{} of sentences name a person, which is why the yes/no question is scored by BAcc, where chance is \YesnoChancePct{} whatever the class balance.

\section{Extended Results}
\label{sec:appendix}
\begin{figure}[t]
\centering
\includegraphics[width=\columnwidth]{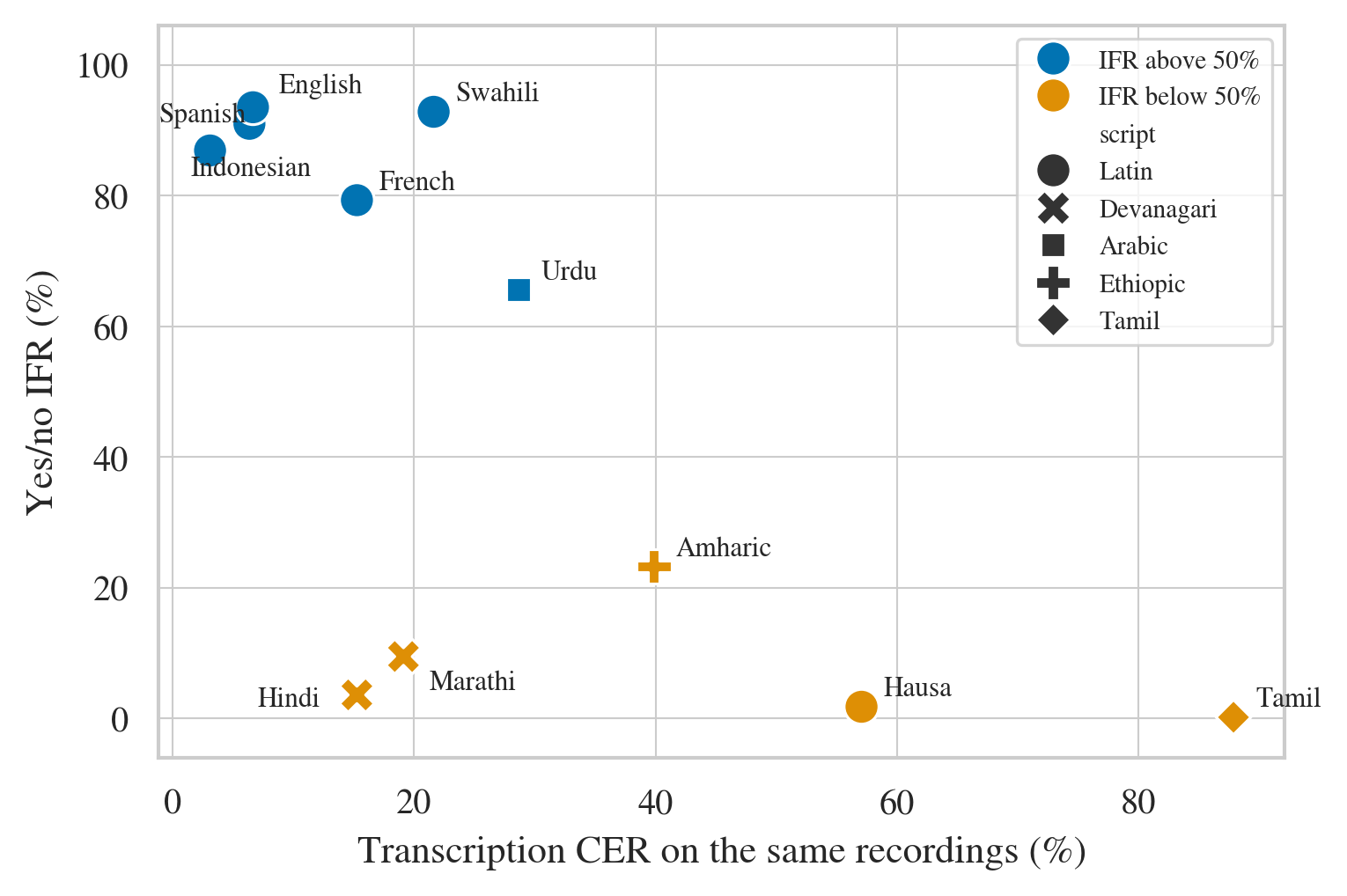}
\caption{Transcription CER against yes/no IFR. Each point is one projector, trained and evaluated on its own language. Marker shape denotes the language's script, and color marks whether its IFR is above \AvailableAtPct{}.}
\label{fig:availability}
\end{figure}
% GENERATED by build_slu_abstract_tables.py -- edit that script, not this file.

\begin{table*}[t]
\centering
\small
\begin{tabular}{lrrrr}
\toprule
System & Acc.\% & Acc.$^\dagger$\% & BAcc\% & IFR\% \\
\midrule
\multicolumn{5}{l}{\emph{Reference points}} \\
Chance (7 classes) & 14.3 & -- & 14.3 & -- \\
Majority-class prediction & 30.6 & -- & 14.3 & -- \\
\midrule
\multicolumn{5}{l}{\emph{Fleurs-SLU \citep{schmidt_fleurs-slu_2025}: a classification head trained on English task data}} \\
MMS-1B encoder & 21.3 & -- & -- & -- \\
SeamlessM4Tv2-L encoder \citep{communication_seamlessm4t_2023} & 80.9 & -- & -- & -- \\
Cascade, LLM2Vec-S4T & 86.4 & -- & -- & -- \\
Reference text, NLLB-LLM2Vec \citep{team_no_2022} & 88.9 & -- & -- & -- \\
\midrule
\multicolumn{5}{l}{\emph{This work: one projector per language behind Tiny Aya Global, no task training}} \\
Digit template & 16.0 & 17.7 & 23.2 & 81.3 \\
In-language template & 10.9 & 13.9 & 20.5 & 65.0 \\
Cascade, projector transcript & 58.3 & 58.3 & 52.5 & 100.0 \\
Reference transcript & 73.5 & 73.5 & 71.5 & 100.0 \\
\midrule
\multicolumn{5}{l}{\emph{Multi-task LALM, our run on the same utterances}} \\
Qwen2-Audio-7B \citep{chu_qwen2-audio_2024} & 37.2 & 37.2 & 39.4 & 100.0 \\
\bottomrule
\end{tabular}
\caption{Topic classification on FLEURS audio across 11 languages. Scores are means over the languages, and balanced accuracy (BAcc) first averages the 7 classes within each language. The Fleurs-SLU rows are that paper\textquotesingle s published results. Its systems are trained on the task, while our projectors and the multi-task LALM are not. The cascade and reference-transcript rows replace the audio with the projector\textquotesingle s transcript or the reference transcript. Acc. counts unanswered utterances as wrong. Acc.$^\dagger$ and BAcc use answered utterances only. Text input is scored at its first named option (Appendix~\ref{sec:metrics}).}
\label{tab:comparison}
\end{table*}

% GENERATED by build_slu_abstract_tables.py -- edit that script, not this file.

\begin{table}[t]
\centering
\scriptsize
\setlength{\tabcolsep}{3pt}
\begin{tabular}{lrrrrrr}
\toprule
 & \multicolumn{2}{c}{Open question} & \multicolumn{2}{c}{In-language} & \multicolumn{2}{c}{Qwen2-Audio-7B} \\
\cmidrule(lr){2-3}\cmidrule(lr){4-5}\cmidrule(lr){6-7}
Language & IFR\% & Acc.$^\dagger$\% & IFR\% & BAcc\% & IFR\% & BAcc\% \\
\midrule
English & 3 & 91 & 100 & 41 & 100 & 67 \\
Spanish & 11 & 68 & 100 & 29 & 100 & 63 \\
French & 1 & 50 & 100 & 34 & 100 & 64 \\
Indonesian & 16 & 48 & 81 & 21 & 100 & 36 \\
Swahili & 1 & 50 & 57 & 14 & 100 & 21 \\
Urdu & 0 & -- & 88 & 14 & 100 & 44 \\
Amharic & 2 & 60 & 67 & 14 & 100 & 31 \\
Hausa & 0 & -- & 68 & 15 & 100 & 17 \\
Hindi & 0 & -- & 13 & 14 & 100 & 45 \\
Marathi & 0 & -- & 8 & 14 & 100 & 27 \\
Tamil & 0 & -- & 33 & 14 & 100 & 19 \\
\bottomrule
\end{tabular}
\caption{Topic classification across languages. Open question task: the projector is prompted for a category word. Accuracy is computed over answered utterances only, and blank where none was answered. In-language prompt: the digit template translated into the projector\textquotesingle s native language. Qwen2-Audio-7B: evaluated on the same utterances using the English digit template. BAcc is balanced accuracy over answered utterances.}
\label{tab:topic-by-language}
\end{table}

% GENERATED by build_slu_abstract_tables.py -- edit that script, not this file.

\begin{table}[t]
\centering
\footnotesize
\setlength{\tabcolsep}{4pt}
\begin{tabular}{lrr}
\toprule
 & \multicolumn{2}{c}{Yes/no, text input} \\
\cmidrule(lr){2-3}
Language & IFR\% & BAcc\% \\
\midrule
English & 89.1 & 52.8 \\
Spanish & 64.7 & 47.2 \\
French & 87.0 & 45.3 \\
Indonesian & 98.5 & 54.4 \\
Swahili & 93.6 & 48.0 \\
Urdu & 58.7 & 50.6 \\
Amharic & 95.3 & 46.8 \\
Hausa & 25.7 & 45.0 \\
Hindi & 95.5 & 46.7 \\
Marathi & 70.8 & 53.5 \\
Tamil & 91.7 & 51.1 \\
\midrule
Mean & 79.1 & 49.2 \\
\bottomrule
\end{tabular}
\caption{The yes/no question with the reference transcript as text input, per language, through Tiny Aya Global. IFR is computed based on the first option named. BAcc is over answered utterances, and chance is \YesnoChancePct{}. The last row is the mean over the \NLangs{} languages.}
\label{tab:yesno-text}
\end{table}

% GENERATED by build_slu_abstract_tables.py -- edit that script, not this file.

\begin{table*}[t]
\centering
\small
\setlength{\tabcolsep}{4pt}
\begin{tabular}{lrrrrrrrrrrrr}
\toprule
Language & \multicolumn{2}{c}{Global} & \multicolumn{2}{c}{Earth} & \multicolumn{2}{c}{Fire} & \multicolumn{2}{c}{Water} & \multicolumn{2}{c}{Base} & \multicolumn{2}{c}{Qwen3-4B} \\
\cmidrule(lr){2-3}\cmidrule(lr){4-5}\cmidrule(lr){6-7}\cmidrule(lr){8-9}\cmidrule(lr){10-11}\cmidrule(lr){12-13}
BAcc\% & S & T & S & T & S & T & S & T & S & T & S & T \\
\midrule
English & 41.4 & 73.0 & 50.6 & 72.0 & 38.2 & 73.8 & 53.4 & 71.5 & 15.6 & 15.1 & 43.6 & 88.0 \\
Spanish & 36.2 & 73.6 & 27.8 & 73.0 & 23.0 & 76.5 & 29.2 & 74.1 & 15.6 & 14.3 & 42.3 & 82.5 \\
French & 35.7 & 75.8 & 37.7 & 76.6 & 62.6 & 78.2 & 48.2 & 75.8 & 15.5 & 17.0 & 56.8 & 84.0 \\
Indonesian & 35.7 & 71.3 & 31.7 & 71.1 & 31.7 & 74.1 & 42.7 & 70.8 & 13.4 & 14.3 & 30.5 & 83.1 \\
Swahili & 17.1 & 66.0 & 22.0 & 68.2 & 18.0 & 73.6 & 19.3 & 65.7 & 16.8 & 14.3 & 18.3 & 62.0 \\
Urdu & 16.4 & 70.9 & 33.7 & 71.7 & 39.6 & 69.4 & 31.3 & 72.3 & 9.3 & 21.0 & 16.3 & 81.3 \\
Amharic & 14.7 & 70.4 & 19.2 & 71.4 & 22.2 & 72.1 & 17.4 & 71.7 & 14.4 & 19.9 & 12.7 & 44.1 \\
Hausa & 14.9 & 67.2 & 20.2 & 65.6 & 12.9 & 72.4 & 14.5 & 65.3 & 16.5 & 14.3 & 16.7 & 42.5 \\
Hindi & 14.3 & 71.5 & 16.7 & 70.9 & 18.1 & 72.6 & 15.5 & 71.4 & 14.3 & 22.2 & 12.4 & 81.6 \\
Marathi & 14.3 & 71.1 & 14.3 & 71.9 & 14.3 & 73.7 & 14.3 & 71.4 & 16.7 & 19.2 & 14.5 & 75.0 \\
Tamil & 14.3 & 75.2 & 14.3 & 76.1 & 14.5 & 75.0 & 14.3 & 75.5 & 14.3 & 20.4 & 16.4 & 82.0 \\
\midrule
Mean & 23.2 & 71.5 & 26.2 & 71.7 & 26.8 & 73.8 & 27.3 & 71.4 & 14.8 & 17.4 & 25.5 & 73.3 \\
\bottomrule
\end{tabular}
\caption{Balanced accuracy (BAcc) over the 7 topics on the digit template question, by language and frozen decoder. Each decoder has its own projector for each language. S uses speech. T uses the reference transcript instead of the audio. The final row averages the \NLangs{} languages. Base is the pretrained Tiny Aya before instruction tuning. Earth, Fire and Water are its regional variants, and Global is reported in the main text. Chance is \TopicChancePct{} (i.e. 1/7).}
\label{tab:decoders}
\end{table*}

\paragraph{Does recognition quality predict instruction following? No.}
Figure~\ref{fig:availability} plots each projector's CER against its yes/no IFR from Table~\ref{tab:perlang}. Hindi and French transcribe equally well, at \PairHiCer{} and \PairFrCer{} CER, yet Hindi answers \PairHiAnsPct{} of the yes/no questions and French \PairFrAnsPct{}. The amount of training data does not predict IFR either. Hindi's projector trains on \HoursHiIn{}~h of WorldSpeech and Urdu's on \HoursUrPk{}~h, yet Urdu's projector answers \UrAnsPct{} of the questions.

\paragraph{Does transcribing first help? Yes.}
In a cascade, the projector first transcribes the audio, and the decoder then answers from that transcript as text. Table~\ref{tab:comparison} lists the cascade beside the audio and the reference transcript. The cascade answers \SluCascadeAnsPct{} of the digit template questions at \SluCascadeRecallPct{} mean BAcc when using the projector's own transcript as text. That is more than twice the \SpeechTopicBalPct{} from the audio directly, and below the \TextTopicBalPct{} from the reference transcript. The projector's transcript contains the topic, but the decoder cannot recover it \emph{directly} from the audio tokens. Further training should therefore target the alignment between the projector's audio tokens and the decoder, which \citet{gopal_language-aware_2026} show can be trained from ASR pairs alone, with the encoder and decoder frozen.

\paragraph{How far is zero-shot prompting from task training?}
Table~\ref{tab:comparison} places our projectors beside the topic classifiers that Fleurs-SLU trains on the task, on our eleven languages. \textbf{Task-trained systems are far ahead.} With a classification head trained on English task data, Fleurs-SLU reaches \SibTextAccPct{} accuracy from the reference text, \SibCascadeAccPct{} through a translation cascade, and \SibSpeechLoPct{}--\SibSpeechHiPct{} from speech encoders. \textbf{Our projectors' raw accuracy stays below the majority-class prediction.} They have no task training and no classification head. Their raw accuracy of \TopicNumAccPct{} sits below the \TopicMajorityPct{} obtained by always predicting the most frequent topic, which is why we report BAcc: \TopicNumBalPct{} against \TopicChancePct{} chance. The multi-task LALM, under the same answer rule, reaches \QwenAudioBalPct{}.

\paragraph{How do the topic results vary across languages?}
Table~\ref{tab:topic-by-language} gives the per-language values behind the topic classification results. \textbf{Open questions are rarely answered.} At most \TopicOpenAnsHiIntPct{} of utterances get a category. When the projector does answer, it is usually right, up to \TopicOpenAccHiPct{}. So the open form measures whether the decoder will commit to answer rather than whether it knows. \textbf{Asking in the projector's own language does not help.} The digit template translated into each projector's language drops IFR to \TopicMlAnsPct{} and accuracy to \TopicMlAccPct{}. It elicits answers less often than the English one in \MlAnsFewerN{} of \NLangs{} languages and scores higher in none. \textbf{The multi-task LALM answers all but one utterance and has higher BAcc than our projectors in every language.} Its raw accuracy still falls below that of always predicting the most frequent topic in \QwenAudioBelowMaj{} of the \NLangs{} languages, \QwenBelowMajSilent{} being languages where our projectors answer few yes/no questions.

\paragraph{Does text input change the yes/no result? No.}
Table~\ref{tab:yesno-text} gives the yes/no results with the reference transcript in place of the audio. The decoder answers \TextYesnoAnsLoPct{}--\TextYesnoAnsHiPct{} of the questions, and its BAcc stays within \TextYesnoBalLoPct{}--\TextYesnoBalHiPct{} in every language, around the \YesnoChancePct{} chance level. The mean of \TextYesnoBalPct{} sits inside the \YesnoBalLoPct{}--\YesnoBalHiPct{} the projectors reach from speech. The yes/no question fails in the decoder, with or without the projector.

\paragraph{Does the decoder matter?}
Table~\ref{tab:decoders} repeats the topic classification setup using the digit template with different frozen decoders: Tiny Aya Global, which the main text reports; its three regional variants Earth, Fire and Water; Tiny Aya Base, the same model before instruction tuning; and Qwen3-4B \citep{yang_qwen3_2025}, a second model family. The projector for each decoder is trained for each language as in the main text. The table answers three questions:
\begin{enumerate}\setlength{\itemsep}{0pt}
\item Does instruction tuning matter? Yes. Without it, the decoder cannot answer.
\item Does the regional variant matter? A little.
\item Does the decoder family matter? Not much. Qwen3-4B performs like Tiny Aya.
\end{enumerate}

\paragraph{1. Instruction tuning.}
The base decoder answers at chance, with \DecBaseSpeechPct{} mean BAcc through speech and \DecBaseTextPct{} from the reference text. Every instruction-tuned decoder reaches \DecTextLoPct{}--\DecTextHiPct{} from the reference text. The base decoder fails even when it reads the text, so no projector can make it answer from speech.

\paragraph{2. Regional variant.}
Through speech, the three regional variants reach \DecRegionalLoPct{}--\DecRegionalHiPct{} mean BAcc, against \TopicNumBalPct{} for Global. Each language is matched to one of the three variants, as listed in Appendix~\ref{sec:training}. Using the matched variant adds \DecMatchGainPp{} points over the mean of the other two.

\paragraph{3. Decoder family.}
Qwen3-4B reaches \DecQwenTextPct{} BAcc from the reference text, inside the range of the instruction-tuned Tiny Aya decoders. Through speech it reaches \DecQwenSpeechPct{}, \DecQwenVsGlobalPp{} points above Global.

\end{document}